\documentclass[aps,prb,superscriptaddress,twocolumn,eprint,amsmath,amssymb,nofootinbib,citeautoscript]{revtex4-2}
\usepackage{amsmath,amsthm,amsfonts,amssymb,amscd}
\usepackage{fullpage}
\usepackage{lastpage}
\usepackage{graphicx}
\usepackage{hyperref}
\usepackage{wrapfig}
\usepackage{enumerate}
\usepackage{fancyhdr}
\usepackage{mathrsfs}
\usepackage{mathtools}
\usepackage{bm}
\usepackage[dvipsnames]{xcolor}
\usepackage{float}
\usepackage[margin=0.55in]{geometry}
\usepackage[english]{babel}

\usepackage{comment}

\usepackage[section]{placeins} 

\newcommand{\beq}{\begin{equation}}
	\newcommand{\eeq}{\end{equation}}

\newcommand{\B}[1]{\textcolor{black}{#1}}

\begin{document}

	\title{Fluctuations can induce local nematic order and extensile stress in monolayers of motile cells
	}
	\author{Farzan Vafa}
	\email[Corresponding author:\ ]{fvafa@ucsb.edu}
	\affiliation{Department of Physics, University of California Santa Barbara, Santa Barbara, CA 93106, USA}
	\author{Mark J. Bowick}
	\affiliation{Kavli Institute for Theoretical Physics, University of California Santa Barbara, Santa Barbara, CA 93106}
	\author{Boris I. Shraiman}
	\affiliation{Department of Physics, University of California Santa Barbara, Santa Barbara, CA 93106, USA}
	\affiliation{Kavli Institute for Theoretical Physics, University of California Santa Barbara, Santa Barbara, CA 93106}
	\author{M.~Cristina Marchetti}
	\email[Corresponding author:\ ]{cmarchetti@ucsb.edu}
	\affiliation{Department of Physics, University of California Santa Barbara, Santa Barbara, CA 93106, USA}

	\date{\today}
	\begin{abstract}
		
		Recent experiments in various cell types have shown that two-dimensional tissues often display local nematic order, with evidence of extensile stresses manifest in the dynamics of topological defects. Using a mesoscopic model where tissue flow is generated by fluctuating traction forces coupled to the nematic order parameter, we show that the resulting tissue dynamics can spontaneously produce local nematic order and an extensile internal stress. A key element of the model is the assumption that in the presence of local nematic alignment, cells preferentially crawl along the nematic axis, resulting in anisotropy of fluctuations. Our work shows that activity can drive either extensile or contractile stresses in tissue, depending on the relative strength of the contractility of the cortical cytoskeleton and tractions by cells on the extracellular matrix. 
		
	\end{abstract}
	
	\maketitle
	
	\section{Introduction}
	
	Nematic order has been widely seen in biological active matter~\cite{Marchetti2013,Doostmohammadi2018,shankar2020topological}, ranging from suspensions of cytoskeletal filaments and associated motor proteins~\cite{Sanchez2012,Keber2014topology,Kumar2018tunable} to cell sheets~\cite{Duclos2016,Saw2017,Kawaguchi2017,Blanch2018}, bacteria collectives~\cite{doostmohammadi2016defect,Nishiguchi2017,copenhagen2020topological} and developing organisms~\cite{maroudas2020topological}. In all of these systems, orientational order is interrupted by topological defects, singular deformations of the order parameter field that cannot be removed by a continuous transformation and provide a fingerprint of the broken symmetry. In two-dimensional nematics, the lowest energy defects are $+1/2$ and $-1/2$ disclinations~\cite{Gennes1993}. A key property of active nematics is that the spontaneous flows induced by activity render the $+1/2$ defects motile~\cite{Narayan2007,Sanchez2012,Giomi2013,thampi2013velocity,Pismen2013dynamics}. Importantly, the direction of motion of the comet-like $+1/2$ defect is controlled by the sign of active stresses: extensile active stresses, as generated in cytoskeletal suspensions by cross-linking motor proteins or in living cells through division, drive the defect to move towards the head of the comet, while contractile stresses, as occurring for instance in the acto-myosin cytoskeleton, drive the defect to move towards the tail~\cite{Giomi2013}. This behavior has been verified extensively in simulations~\cite{Doostmohammadi2018}. A surprising experimental finding is that in almost all realizations of two-dimensional active nematics, the motion of the $+1/2$ defect suggests that extensile activity dominates. While in bacteria this has been clearly associated with cell division~\cite{Doostmohammadi2015,DellArciprete2018a}, the origin of the extensile activity  observed in layers of human bronchial epithelial cells~\cite{Blanch2018}, in Madine-Darby canine kidney (MDCK) cells~\cite{Saw2017}, in stem cells~\cite{Kawaguchi2017}, and in simulations~\cite{Mueller2019} on times scales shorter than those associated with appreciable cell division \B{has only recently been explored~\cite{balasubramaniam2020nature}}.
	
	The dynamics of cells crawling on a substrate is generally driven by two types of active mechanisms: (i) internal active forces generated within a cell  by myosin-driven cellular contractility and transmitted throughout the tissue via supracellular actin coordination,  and (ii) traction forces between the tissue and the substrate that drive cell motility.  Internal forces due to cellular contractility  yield active stresses, but no net force, while tractions provide an external forcing of the tissue. A naive expectation is that contractile active stresses dominate  the behavior of epithelial tissue, while active tractions provide the main contribution to  cell dynamics in mesenchymal tissue. The relative role, though, of these two types of active processes remains to be quantified.  Here we use a continuum model of an incompressible tissue on a substrate to examine the interplay between these two types of activity. We show that \emph{fluctuating} traction forces advected by flow can generate both local nematic order and effective extensile stresses in an otherwise isotropic tissue.   Essentially, polar active migration drives local cell alignment captured by a nematic order parameter. A key assumption of our model is that cells will preferentially move along the direction of local cell alignment, rather than transverse to it. This is incorporated in the model by coupling the  noise that drives the fluctuating cell tractions to local nematic order.   By working perturbatively in the noise strength, we show that the nonlinear advection of active tractions by cellular flow can be recast in the form of a mean active force that has the structure of an extensile stress and drives cell dynamics. Flow alignment generated by this active forcing in turn destabilizes the isotropic state, suggesting the onset of a state with spatially modulated texture and rotating director field. While the fact that extensile activity can build local nematic order in what would be an isotropic state in the passive limit has been highlighted before~\cite{srivastava2016negative,putzig2016instabilities,santhosh2020activity}, a new result of our model is that the extensile stress observed in crawling tissue may arise from fluctuating cellular tractions. 
	
	The organization of this paper is as follows. In Sec.~\ref{model}, we introduce the model.  In Sec.~\ref{computation}, we evaluate the mean force from fluctuating tractions and show that it yields effective extensile stresses that generate local nematic order. We conclude in Sec.~\ref{discussion} with a brief discussion.
	
	\section{Model of Crawling Tissue }
	\label{model}
	
	We describe the tissue at the continuum level \B{as a contractile active gel~\cite{prost2015active}} in terms of the cell density $\rho$, the flow velocity $\mathbf{v}$ and a nematic order parameter $ {\mathbf{Q}} = (2\mathbf{\hat n}\mathbf{ \hat n} - \mathbf{1})S$ that captures cell shape anisotropy and alignment of elongated cells, with $S$  the magnitude of  nematic order and   $\mathbf{\hat n}$ the unit nematic director field that identifies the direction of local order. \B{The tissue is in contact with a substrate and force balance requires that the force on a tissue element due to the surrounding cells be balanced by the traction force density $\mathbf{t}$ that the tissue exerts of the substrate, according to
		\begin{equation}
			\partial_j\sigma_{ij}-t_i=0\;,
			\label{eq:balance}
		\end{equation}
		where $\sigma_{ij}=-p\delta_{ij}+2\eta S_{ij}+\sigma_{ij}^a$ is the tissue stress tensor, with $p$ the pressure, $\eta$ the tissue shear viscosity, $S_{ij} = \frac{1}{2}(\partial_i v_j + \partial_j v_i)$ the rate of strain tensor, and $\sigma_{ij}^a=\alpha_cQ_{ij}$ an active stress. The traction force is written as $\mathbf{t}=\Gamma\mathbf{v}-\mathbf{f}$, where $\Gamma$ is the effective friction with the substrate and
		$\mathbf{f}$  a \emph{fluctuating}  propulsive force density caused by transiently polarized cryptic lamellipodia activity which underlies cell motility.
		This form for the traction $\mathbf{t}$ allows for local misalignment
		of cell propulsion and tissue velocity due to intracellular interactions, consistent
		with experimental findings~\cite{brugues2014forces,notbohm2016cellular}.} Assuming the tissue to be incompressible ($\bm\nabla\cdot\mathbf{v}=0$, hence $\rho={\rm constant}$), the dynamics is  described by the following equations
	\begin{gather}
		\eta \nabla^2 \mathbf{ v }- \bm\nabla p + \alpha_c\bm\nabla\cdot\mathbf{Q}=\Gamma\mathbf{v} - \mathbf{ f} \label{v} \; ,\\
		D_t \mathbf{Q} = \lambda \mathbf{S}  - \frac{1}{\gamma}\left[\frac{\delta F( {\mathbf{Q}})}{\delta  {\mathbf{Q}}}\right]^{ST}
		\label{Q}\;,\\
		\left(\partial_t+\mathbf{v}\cdot\bm\nabla\right)\mathbf{f} = - \frac{\mathbf{f}}{\tau}  + \bm\xi \; , \label{f} 
	\end{gather}
	where $D_t = \partial_t + \mathbf{v}\cdot\bm\nabla - [\bm\Omega,\cdot]$ is the comoving and corotational derivative, \B{with $\Omega_{ij} = \frac{1}{2}(\partial_i v_j - \partial_j v_i)$ the vorticity tensor,} and the superscript $ST$ in Eq.~\eqref{Q} denotes the symmetric traceless part of any tensor
	~\cite{Marchetti2013,Doostmohammadi2018}. 
	\B{Other couplings to flow gradient are generally allowed in Eq.~\eqref{f} for $f_i$, such as  $\Omega_{ij}f_j$ and $\lambda S_{ij}f_j$. These terms, however, do not contribute to the noise-renormalized mean propulsion force evaluated below and are therefore not included in the equation.} 
	
	The Stokes equation, Eq.~\eqref{v}, includes two types of active processes. First, force dipoles due to the pulling action of myosins and transmitted across the epithelial tissue by cell-cell interactions mediated by E-cadherins result in \B{an \emph{apolar}} contractile active stresses $\alpha_c\mathbf{Q}$~\cite{Simha2002,Marchetti2013,prost2015active}, with $\alpha_c>0$ an activity parameter that incorporates the biochemical processes responsible for cellular contraction and controlled by myosin density and ATP concentration. Second, \B{the presence of the substrate allows for \emph{polar} terms described by the \emph{fluctuating}} propulsive force density $\mathbf{f}$. \B{This may arise,   for instance, from intermittent protrusions and retractions of cryptic lamellipodia at a rate controlled by cell-substrate interaction mediated by focal adhesion complexes. We assume that  the traction force density $\mathbf{f}$ tends to align with the long axis of the cell that controls the direction of local nematic order, but switches direction on a time scale $\tau$  much shorter than the time scale $\tau_Q$ controlling the reorientation of the local nematic texture. As a result, there is no polar order of propulsive forces at the tissue scale. This separation of time scales allows us to treat this source of activity independently and examine how microscopic cell scale fluctuations feed back  on the tissue-scale active stress.  } \B{ Finally, we have neglected in the Stokes equation elastic liquid crystalline stresses. These terms are of higher order in the gradients than the active stress and do not change the results described below. }
	
	The dynamics of the nematic order parameter is controlled by  flow alignment driven by coupling to vorticity and rate of strain, with $\lambda$ the flow alignment parameter, and relaxation controlled by the de Gennes-Ginzburg-Landau free energy~\cite{Landau1998, Gennes1971},
	$F=\frac{1}{2}\int_{\mathbf{x}}\left\{a ~Tr[\mathbf{Q}^2]+K(\bm\nabla\cdot\mathbf{Q})^2\right\}$, with $K$ the nematic stiffness (in the one-elastic constant approximation) and $\gamma$ the rotational friction of the nematic. We assume that the tissue is isotropic ($a>0$), and hence neglect stabilizing terms of order $\mathcal{O}(\mathbf{Q}^4)$ in the free energy, so that 
	\beq
	-\frac{1}{\gamma}\frac{\delta F( {\mathbf{Q}})}{\delta  {\mathbf{Q}}}=-\frac{\mathbf{Q}}{\tau_Q}+D\nabla^2\mathbf{Q}\;,
	\label{relax}
	\eeq
	with $\tau_Q=\gamma/a$ the relaxation time of the nematic texture and $D=K/\gamma$ a diffusivity.
	
	\B{The fluctuating local propulsive force $\mathbf{f}$ (Eq.~\eqref{f}) persists over a time $\tau$, and is then randomized by interactions with other cells and short scale active processes embodied by 
		a zero mean  random force $\bm\xi$,} with correlations
	\begin{gather}
		\B{\langle}\xi_i(\mathbf{x},t)\xi_j(\mathbf{x}',t')\B{\rangle} =  \varepsilon^2 \delta(t - t')\delta(\mathbf{x}-\mathbf{x}')[\delta_{ij} + \kappa \langle Q_{ij}(\mathbf{x},t)\rangle] \; ,
		\label{std}
	\end{gather}
	with $\kappa>0$ and $\varepsilon^2$ the strength of the noise.   
	The noise correlations are chosen to depend on $\mathbf{Q}$  so as to capture the fact that local alignment will result in different cell motility along and transverse to the director. The positive sign of $\kappa$ favors motion along the director and penalizes displacements transverse to the direction of local order. We also  assume that $\kappa S < 1$ to ensure that the variance of the noise is positive.  
	\B{Given $\tau\ll\tau_{Q}$,  we then examine 
		the  behavior of the tissue on times long compared to $\tau$, but in the presence of finite local nematic order.  The separation of these two time scales in epithelial tissue is evidenced by the observation of negligible cell motility (corresponding to $\langle\mathbf{f}\rangle=0$) and appreciable regions of local nematic order (corresponding to  $\langle\mathbf{Q}\rangle\not=0$). }

	To impose incompressibility we take the divergence of Eq.~\eqref{v} to obtain
	$\nabla^2 p = \bm\nabla\cdot \left(\mathbf{f}+\alpha_c\bm\nabla\cdot\mathbf{Q}\right)$. 
	Eliminating the pressure, the Stokes equation can  then be formally written as
	\beq
	\Gamma v_i - \eta \nabla^2 v_i = \mathcal P_{ij}\left(f_j + \alpha_c\partial_lQ_{jl}\right)\;, \label{vReduced}
	\eeq
	with
	$\mathcal P_{ij}= \left(\delta_{ij} - \nabla^{-2}{\partial_i \partial_j}\right)$
	being the incompressibility projection operator.

	\section{Noise induced extensile stress and nematic order}
	\label{computation}
	
	To linear order, the only steady state solution of our noise-averaged equation is $\langle\mathbf{v}\rangle=\langle\mathbf{f}\rangle=\langle\mathbf{Q}\rangle=0$.
	In this section we show, however, that noisy traction forces renormalize the flow velocity, inducing both nematic alignment and extensile active stresses.
	
	Taking the Fourier transform in space and time, Eqs.~\eqref{vReduced} and \eqref{f} can be written as 
	\begin{align}
		v_i({\bf k},\omega) &= \frac{1}{\Gamma + \eta k^2} \left(\delta_{ij} - \hat{k}_i\hat{k}_j\right)\left[f_j({\bf k},\omega) +\alpha_cik_lQ_{jl}({\bf k},\omega)\right] \;,\label{eq:vFT}\\
		f_i({\bf k},\omega) &= \frac{\xi_i({\bf k},\omega)}{\tau^{-1} + i\omega} \nonumber\\
		&- \frac{i k_j}{\tau^{-1} + i\omega} \int_{\omega'} \int_{\bf q}v_j({\bf q},\omega')f_i({\bf k-q},\omega - \omega') \; , \label{eq:fFT}
	\end{align}
	where $\int_{\omega}=\int \frac{d\omega}{2\pi}$ and $\int_{\mathbf{q}}= \int \frac{d{\bf q}}{(2\pi)^2}$.
	Substituting Eq.~\eqref{eq:vFT} into Eq.~\eqref{eq:fFT}, and \B{using the fact that due to the separation of time scales we can treat $\mathbf{f}$ and the noise $\bm\xi$} as  uncorrelated with  $\mathbf{Q}$, we calculate the renormalization of the traction force to first order in the noise amplitude, with the result 
	\begin{align}
		\langle f_i({\bf k},\omega)\rangle = &{-\frac{\varepsilon^2\kappa\ln(\ell_v/a)}{4\pi\eta}\frac{ik_j\langle Q_{ij}({\bf k},\omega)\rangle}{(\tau^{-1} + i\omega)(2\tau^{-1}+ i\omega)}}  \; ,
		\label{meanf}
	\end{align}
	where we have used
	\beq
	\int_{\mathbf{q}}\frac{\delta_{ij}-\hat{q}_i\hat{q_j}}{\Gamma+\eta q^2}=\frac{\delta_{ij}}{8\pi\eta}\ln\left(1+\frac{\ell_v^2}{a^2}\right)\approx
	\frac{\delta_{ij}}{4\pi\eta}\ln\left(\frac{\ell_v}{a}\right) \;,
	\eeq
	with $\ell_v=\sqrt{\eta/\Gamma}$ the viscous screening length and $a$ a short scale cutoff of the order of the cell size.
	
	Since we are interested in time $t \gg \tau$, with $\tau\ll\tau_Q$, we can neglect $\omega$ in the denominator of Eq.~\eqref{meanf}. 
	Taking the inverse Fourier transform yields a mean force
	\beq 
	\langle\mathbf{ f}\rangle = \alpha_f\bm\nabla\cdot  \langle\mathbf{Q}\rangle \; ,
	\label{f_final}
	\eeq
	where 
	\beq\alpha_f = -\frac{\varepsilon^2\kappa\tau^2}{8\pi\eta}
	\ln\left(\frac{\ell_v}{a}\right)<0 \; , \label{eq:alpha_f}\eeq
	is an extensile activity.
	\B{Extensile stresses  arise because persistent cell tractions $\mathbf{f}$ along the direction of local nematic order tends to elongate local regions of the tissue in that direction. 
		This effect is transient (only lasts a short time $\tau$), but due to the nonlinearity of the advection term in the $\mathbf{f}$ equation it leads to a nonzero value of $\langle \mathbf{f}\rangle$ that corresponds to extensile stresses.}\footnote{\B{This is seen by writing $\partial_t f_i\sim - (\mathbf{v}\cdot\bm\nabla)f_i\sim -(f_j\nabla_j)f_i$, and so $\langle f_i\rangle \sim -\langle\tau (f_j\nabla_j)f_i\rangle\sim -(...)\nabla_j\langle Q_{ij}\rangle$.}}.
	
	We now return to Eqs.~\eqref{vReduced} and \eqref{Q}, average over noise, and use Eq.~\eqref{f_final} to eliminate the mean traction force. The linear dynamics of fluctuations from the quiescent disordered state with $\langle\mathbf{v}\rangle=\langle\mathbf{Q}\rangle = 0$ is then governed by the equations
	\begin{gather}
		\Gamma  \langle v_i\rangle - \eta \nabla^2 \langle v_i\rangle = \alpha\mathcal P_{ij} \partial_l\langle Q_{jl}\rangle\;, \label{vtot}\\
		\partial_t\langle Q_{ij}\rangle = \frac{\lambda}{2} \left(\partial_i\langle v_j \rangle +\partial_j\langle v_i \rangle \right)- \frac{\langle Q_{ij}\rangle}{\tau_Q}+D\nabla^2\langle Q_{ij}\rangle\;,
		\label{Qtot}
	\end{gather}
	where 
	\beq
	\alpha=\alpha_c +\alpha_f
	\eeq
	is the total activity, with sign controlled by the interplay of contractile activity ($\alpha_c>0$) from actomyosin contractility and extensile activity ($\alpha_f<0$) from fluctuating propulsive forces.
	
	We next demonstrate that extensile stresses from fluctuating tractions also build up local nematic order, justifying our choice of an anisotropic noise correlator.  Taking the spatial Fourier transform of Eqs.~\eqref{vtot} and \eqref{Qtot} and eliminating the velocity in favor of the alignment tensor, we obtain a closed equation for $\langle Q_{ij}(\mathbf{k},t)\rangle$, similar to~\cite{srivastava2016negative,putzig2016instabilities}:
	\begin{align}
		\partial_t \langle Q_{ij}\rangle &= {-\left(\tau_Q^{-1}+Dk^2\right) \langle Q_{ij}\rangle}\;,\notag\\
		&+\frac{\alpha\lambda}{2(\Gamma + \eta k^2)}\left[-k_ik_l \langle Q_{jl}\rangle-k_jk_l\langle Q_{il}\rangle + 2k_ik_j\Psi^\parallel \right] 
		\; ,
		\label{eq:deltaQDyn}
	\end{align}
	where $\Psi^\parallel = \hat{k}_i\hat{k}_j\langle Q_{ij}\rangle$. Upon contraction of Eq.~\eqref{eq:deltaQDyn} with $\hat{k}_i\hat{k}_j$ and $\epsilon_{is}\hat{k}_s\hat{k}_j$, the longitudinal mode $\Psi^\parallel$ and the transverse mode $\Psi^\perp = \epsilon_{is}\hat{k}_s\hat{k}_j \langle Q_{ij}\rangle$  decouple, giving 
	\begin{align}
		\partial_t \Psi^\parallel(\mathbf{k},t) &= -\left(\tau_Q^{-1}+Dk^2\right)\Psi^\parallel (\mathbf{k},t)\;,\\
		\partial_t \Psi^\perp(\mathbf{k},t) &= -\left[\tau_Q^{-1}+Dk^2+\frac{\alpha\lambda k^2/\Gamma}{1 + \ell_v^2 k^2}\right]\Psi^\perp(\mathbf{k},t)\;.
	\end{align}
	The decoupling of longitudinal and transverse modes follows from isotropy. Clearly $\Psi^\parallel$ is always stable. On the other hand, the mode controlling the decay of $\Psi^\perp$ can change sign  if $\alpha<0$, corresponding to the case where extensile activity exceeds contractile activity.  The homogeneous isotropic state becomes unstable for 
	$\alpha<-\alpha^*$, with 
	\beq
	\alpha^*=\frac{\Gamma D}{\lambda}\left(1+\sqrt{\frac{\eta}{\Gamma D\tau_Q}}\right)^2\;.
	\eeq
	
	\begin{figure}[t]
		\includegraphics[width=\columnwidth]{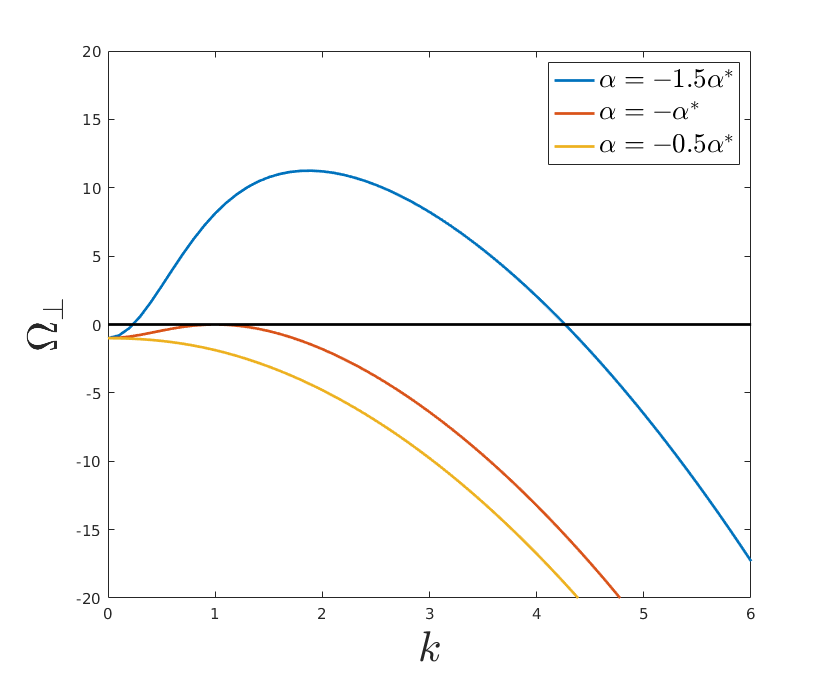}
		\includegraphics[width=\columnwidth]{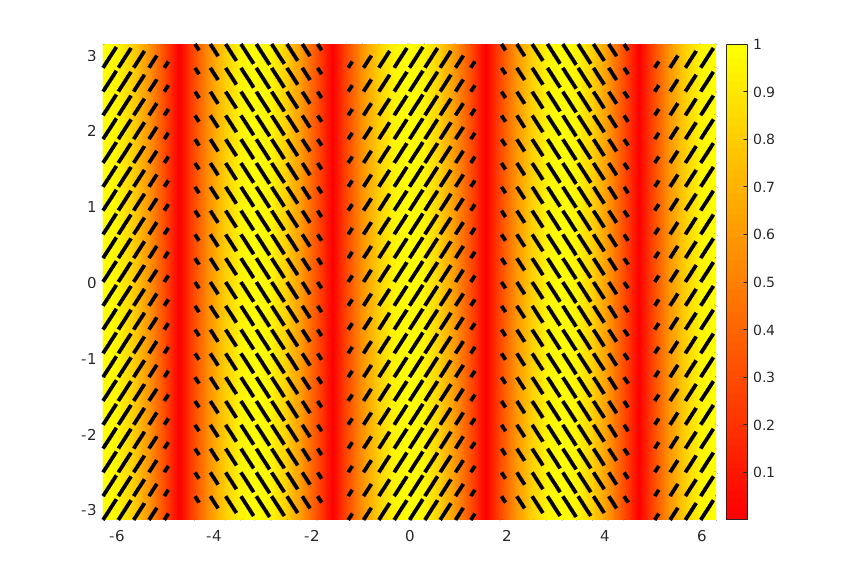}
		\caption{Top:  the dispersion relation of the growth rate $\Omega_\perp$ (Eq.~\eqref{eq:Omega}) as a function of the wavenumber $k$ for three values of  activity: $\alpha = -0.5\alpha^*, \alpha = -\alpha^*$, and $\alpha = -1.5\alpha^*$. At the onset of instability for $|\alpha|=\alpha^*$, only the mode $k = k_0$ is unstable. Above the transition, the unstable modes lie in a band $k_- < k < k_+$. Bottom: sketch of the nematic texture corresponding to the unstable mode with $\mathbf{k} = k_0\mathbf{\hat x}$ for $\phi=0$. The length of the segments, as well as the color, is proportional to the strength of nematic order $S$.
			The angle between the
			director and the wavevector $\mathbf{k}$ is $\pm\pi/4$.}
		\label{fig:instability}
	\end{figure}
	
	\B{This linear instability discussed here for the case of rod-like ($\lambda>0$) extensile ($\alpha<0$) active entities also occurs for disc-like   ($\lambda<0$) contractile ($\alpha>0$) systems, similar to the effect noted in a different context in Ref.~\cite{giomi2010sheared}.} It is best discussed in terms of three length scales that control the dynamics of our model of cellular active nematics: (i) the viscous length $\ell_v$, (ii) the nematic correlation length $\xi=\sqrt{D\tau_Q}$,  and (iii) an active length $\ell_a=\sqrt{|\alpha|\lambda\tau_Q/\Gamma}$ that balances flows induced by active stresses against frictional dissipation~\footnote{The active length defined here is distinct from the one commonly used in the literature, $\ell_{\textit{active}}=\sqrt{K/|\alpha|}$, which controls the dynamics of active nematics in the absence of substrate friction~\cite{Doostmohammadi2018,hemingway2016correlation}.}. In terms of these length scales, the dispersion relation of the mode controlling the dynamics of $\Psi^\perp$  is given by
	\beq 
	\Omega_\perp = -\tau_Q^{-1}\left[1+\xi^2k^2+\rm{sgn}(\alpha)\frac{\ell_a^2k^2}{1 + \ell_v^2 k^2}\right]\;. \label{eq:Omega}
	\eeq

	This mode becomes unstable  ($\Omega_\perp>0$) for extensile systems ($\alpha<0$) of elongated particles ($\lambda>0$) provided
	\beq
	\ell_a >\xi+\ell_v \;.
	\eeq
	The dispersion relation for a few values of parameters is shown in Fig.~\ref{fig:instability}. The wavevector of the fastest growing mode is  $k_0=\frac{1}{\ell_v}\left(\frac{\ell_a}{\xi}-1\right)^{1/2}$, and the instability occurs in a band of wavevectors given by
	\beq
	k_\pm^2(\alpha) = \frac{\ell_a^2 - \xi^2 - \ell_v^2 \pm \sqrt{(\ell_a^2 - \xi^2 - \ell_v^2)^2 - 4\xi^2\ell_v^2}}{2\xi^2\ell_v^2}\;,
	\eeq
	with $k_\pm(|\alpha|=\alpha^*)=k_0$.

	The instability of the disordered state indicates that in extensile systems of elongated units, active flows build up local nematic order, as previously demonstrated for compressible nematics~\cite{srivastava2016negative,putzig2016instabilities}. Note that a finite viscosity is required to stabilize the system at short scales. The instability corresponds to  growth of $\Psi^\perp$, and the associated nematic texture is obtained as solution of  $\hat{k}_i\hat{k}_i\langle\tilde{Q}_{ij}\rangle=0$. The component of the texture corresponding to wavevector $\mathbf{k}$  is then given by
	\beq 
	Q_{ij}(\mathbf{x})= A\cos(\mathbf{k} \cdot \mathbf{x} + \phi) (\epsilon_{i\ell}\hat k_\ell \hat k_j + \epsilon_{j\ell} \hat k_\ell \hat k_i)\;,
	\label{eq:Qk}
	\eeq
	where we have explicitly included the Fourier factor $\cos(\mathbf{k} \cdot \mathbf{x} + \phi)$. This can be written  in terms of a unit nematic director field  $\mathbf{\hat n}$ as ${\mathbf{Q}} = (2\mathbf{\hat n}\mathbf{ \hat n} - \mathbf{1})S$, where  $\mathbf{\hat n}$  is  the eigenvector of the largest eigenvalue of the matrix texture given in Eq.~\eqref{eq:Qk},
	\beq \hat n_i = \frac{1}{\sqrt{2}}[\delta_{ij} - \rm{sgn}[\cos(\mathbf{k} \cdot \mathbf{x} + \phi)]\epsilon_{ij}]\hat k_j \; .
	\eeq
	This satisfies $\cos(\mathbf{\hat{k}}\cdot\mathbf{\hat{n}})=1/\sqrt{2}$ and so $\mathbf{\hat n}$ is at an angle $\pm \pi/4$ to   $\bf{k}$. A sketch of $\mathbf{\hat n}$ for the mode $\mathbf k = k_0 \mathbf{\hat x}$ with amplitude modulated by $A|\cos (k_0x+\phi)|$ is shown in Fig.~\ref{fig:instability}. It corresponds to a modulated chevron texture of periodicity $2\pi/k_0$, with alternating nematic domains tilted at $90^\circ$ to each other, separated by thin isotropic bands. The structure is analogous to that obtained in certain passive nematic liquid crystals under shear~\cite{calderer1996chevron,baza2020shear} and  in lamellar phases of diblock copolymers~\cite{gido1994lamellar}. 
	
	\section{Discussion}
	\label{discussion}

	Using a mesoscopic model for a tissue layer, we have shown that fluctuations in the traction forces exerted by cells on the substrate can build up local nematic order in the tissue and lead to extensile active stresses that compete with those arising from actomyosin contractility.  The sign of the net activity, $\alpha=\alpha_c+\alpha_f$, with $\alpha_c>0$ controlled by contractile actomyosin forces and $\alpha_f<0$ (given in Eq.~\eqref{eq:alpha_f}) determined by the interplay of single-cell motility and global tissue flow, is determined by the larger of these two contributions.
	The tissue dynamics may exhibit overall contractile or extensile behavior depending on the relative magnitude of these two contributions to the active stress.
	In confluent tissues with strong extracellular actin fibers capable of transmitting stresses across cells, we expect a dominance of actomyosin contractility. In mesenchymal tissue, in contrast, local tractions may dominate and mediate the build-up of local nematic order, with associated extensile dynamics of topological defects. 
	Other mechanisms not directly considered here may also enable force transmission across the tissue. In particular, passive cell-cell adhesion  provides intercellular couplings  which enable one cell to \emph{pull} another. Some of these effects are encoded in the tissue shear viscosity $\eta$.  Finally, although here we have assumed incompressibility, which results in  cells pushing on each other, finite compressibility may arise in epithelial cells due, for instance, to deformation of the nucleus or from apical surface tension. The role of the resulting density fluctuations remains to be explored. 
	
	Our work highlights the key role of noise arising from subcellular active processes in mesoscopic models of tissue, something that has been little explored, and opens up several future directions. First, experiments in epithelia~\cite{Saw2017}, stem cells~\cite{Kawaguchi2017}, and \emph{M. xanthus}~\cite{copenhagen2020topological} have  observed cell accumulation near the core of the $+1/2$ defect and cell depletion near the core of the $-1/2$ defect. A valuable extension of our work would be to examine these effects by allowing for density variations and finite tissue compressibility.
	Second, in many situations, cells are also capable of coordinating their motion, leading to emergent or persistent migration. Well-known examples are the collective directed migrations of follicular cells in the \emph{Drosophila} egg-chamber~\cite{Montell2008morphogenetic,barlan2017fat2} and of epithelial cells in wound healing~\cite{Poujade2007collective,Hakim2017collective}. In cancer progression metastatic cells can additionally regulate their dynamical state and switch between collective and single-cell migration~\cite{friedl2017tuning}. The emergence of collective motility has been modeled in the literature by assuming local alignment  of traction forces  of neighboring cells or alignment of cell tractions with the local tissue flow.  At the continuum level, these types of interactions yield an instability associated with a change  in the sign of the traction damping rate $\tau^{-1}$  plus a saturating cubic term, as can be derived from mesoscopic Vicsek-type models~\cite{vicsek1995novel,Bertin2009hydrodynamic,Toner2005hydrodynamics}. It would be interesting to explore whether effective alignment and collective migration can emerge from noisy tractions by, say, including in the traction dynamics a coupling to velocity with ``memory'' of recent flows. This is, however, left for future work.

	\section*{Conflicts of interest}
	
	There are no conflicts of interest to declare.
	
	\acknowledgments
	
	We thank Zvonimir Dogic, Sebastian Streichan and Zhihong You for valuable discussions.
	The work was supported by the NSF through grants PHY-1748958 (MJB), DMR-2041459 (MCM, FV), DMR-1720256 (iSuperSeed) (MCM, MJB) and PHY-0844989 (BIS).
	
	\bibliography{refs}

\end{document}